\documentclass[12pt]{article}
\usepackage[dvips]{color}
\usepackage{epsfig}
\usepackage{amsmath}
\usepackage{graphicx}

\begin{document}
\title{\bf An effective quintessence field with a power-law potential}
\author{\small {M. Khurshudyan$^{a,}$ \thanks{Email:
khurshudyan@yandex.ru}, B. Pourhassan$^{b,}$\thanks{Email: bpourhassan@yahoo.com}, R. Myrzakulov$^{c,}$ \thanks{Email:rmyrzakulov@gmail.com}, S. Chattopadhyay$^{d,}$ \thanks{Email:surajcha@iucaa.ernet.in, surajitchatto@outlook.com}}\\
$^{a}${\small {\em Department of Theoretical Physics, Yerevan State
University,}}\\
{\small {1 Alex Manookian, 0025, Yerevan, Armenia}}\\
$^{b}${\small {\em Department of Physics, Damghan University,
Damghan, Iran}}\\
$^{c}${\small {\em Eurasian International Center for Theoretical Physics, Eurasian National University,}}\\
{\small {Astana 010008, Kazakhstan}}\\
$^{d}${\small {\em Pailan College of Management and Technology, Bengal Pailan Park, Kolkata-700 104, India}}}  \maketitle
\begin{abstract}
In this paper, we consider an effective quintessence scalar field with a power-law potential interacting with a $P_{b}=\xi q\rho_{b}$ barotropic fluid as a first model, where $q$ is a deceleration parameter. For the second model we assume viscous polytropic gas interacting with the scalar field. We investigate problem numerically and analyze behavior of different cosmological parameter concerning to components and behavior of Universe. We also compare our results with observational data to fix parameters of the models. We find some instabilities in the first model which may disappear in the second model for the appropriate parameters. Therefore, we can propose interacting quintessence dark energy with viscous polytropic gas as a successful model to describe Universe.
{\bf Pacs Number(s):} 95.35.+d, 95.85.-e, 98.80.-k.
\end{abstract}

\section{\large{Introduction}}
Accelerated expansion of Universe observed and confirmed \cite{Riess}-\cite{Amanullah}. According to the most observational data such as the Wilkinson Microwave Anisotropy Probe (WMAP) the dark energy (DE), which is characterized by negative pressure, includes most of the energy of our Universe. Extensive reviews on DE include \cite{DE1,DE2,DE3,DE4,DE5,DE6,DE7}. The rest of Universe include ordinary matter and forms only about $4\%$ of the Universe. Today, the origin, dynamics and other several questions concerning to dark components are great challenges in theoretical cosmology.
In general relativity, concepts of dark energy and dark matter were introduce by hand to explain accelerated expansion of the Universe reported by several analysis. However, nature of dark energy and dark mater as well as about possible interactions etc are still open questions of the field. Dark energy thought to be responsible to accelerated expansion with negative pressure strong enough to play the role of antigravity. On theoretical and phenomenological levels, scalar fields were considered as a base of dark energy. Some of them are Tachyonic and quintessence scalar fields. In theoretical cosmology, dynamical models of dark energy, modification of the geometrical part of gravitational action, fluids with EoS equation of more general form $F(\rho,P)=0$  were proposed. Interaction between cosmic fluids also is a topic of great interests giving different modifications of the interaction term.\\
Among different attempts to explain accelerated expansion of the Universe different scalar fields were associated with dark energy. One of them is Tachyonic field with its relativistic Lagrangian,
\begin{equation}\label{eq:tach lag}
L_{TF}=-V(\phi)\sqrt{1-\partial_{i}\phi\partial^{j}\phi},
\end{equation}
which captured a lot of attention (see, for instance, references in \cite{Murli}, \cite{Khurshudyan-Jafar} and \cite{Khurshudyan-Jafar2}).
The stress energy tensor,
\begin{equation}\label{eq:energy tensor}
T^{ij}=\frac{\partial L}{\partial (\partial_{i}\phi)}\partial^{k}\phi-g^{ik}L,
\end{equation}
gives the energy density and pressure as,
\begin{equation}\label{eq:tachyonic density}
\rho=\frac{V(\phi)}{\sqrt{1- \partial_{i}\phi \partial^{i}\phi}},
\end{equation}
and,
\begin{equation}\label{eq:tachyonic pressure}
P=-V(\phi)\sqrt{1- \partial_{i}\phi \partial^{i}\phi}.
\end{equation}
A quintessence field [14, 15] with possibility of interaction [16] are other models based on scalar field with standard kinetic term. Moreover there are other models of scalar field such as phantom model [17], quintom model [18], or k-essense model [19, 20].\\
There are also interesting models to describe dark energy based on is Chaplygin gas (CG) equation of
state [21, 22]. However, ordinary CG model is not consistent with observational data. So, generalized
Chaplygin gas (GCG) model proposed [23], with ability of unification of both dark matter and dark energy.
It is also interesting to study possibility of viscosity in GCG [24-29]. However, observational data ruled out
such a proposal, and the modified Chaplygin gas (MCG) model introduced [30]. Recently, viscous MCG is
also suggested and studied [31, 32]. A further extension of CG model is called modified cosmic Chaplygin
gas (MCCG) which was proposed recently [33-36]. Also, various Chaplygin gas models were studied from
the holography point of view [37-39]. Many recently, extended Chaplygin gas cosmology also introduced to obtain comprehensive model of Chaplygin gas [40].\\
Therefore one can construct a cosmological model based on both vacuum energy and cold dark matter (CDM)as $\Lambda$CDM model. Also, quintessence scalar field $\phi$ with the equation of state $-1 < \omega < 0$, and potential $V(\phi)$ may describe the energy density and the negative pressure. We know that there are two problems arising from all of above. These problems are the fine-tuning, and cosmic coincidence problems. The cosmic coincidence problem may be solved by consideration of a coupling between quintessence DE and CDM, which may considered as some functions of the scalar field $\phi$. It has been shown that if we assume the quintessence scalar field $\phi$ evolves in an exponential potential and let the CDM particle mass also depend exponentially on $\phi$, the late time
behavior of the cosmological equations gives accelerated expansion and, a constant ratio between the dark matter density $\rho_{DM}$ and the dark energy density $\rho_{DE}$. Therefore, interacting quintessence can solves the cosmic coincidence problem [41-47].\\
Our motivation to do the present research is construction of a comprehensive cosmological model in agreement with some observational data which will be used as a toy model to describe Universe.
The paper organized as follow: in the next section we will introduce our models and then in section 3 we write the equations which governs our model. In section 4 we present corresponding result of non-interacting model based on barotropic fluid with the deceleration parameter as a linear function. In section 5 we discuss about interacting models and in section 6 observational constraints investigated to fix parameters of the model. Finally in section 7 we give conclusion and outline for future works.
\section{\large{Models}}
In general sense, usually, three forms of interaction or coupling $Q$ between DE and DM are used,
\begin{equation}\label{eq:Q1}
Q=3Hb\rho_{d},
\end{equation}
\begin{equation}\label{eq:Q2}
Q=3Hb(\rho_{d}+\rho_{m}),
\end{equation}
and,
\begin{equation}\label{eq:Q3}
Q=3Hb\rho_{m},
\end{equation}
where $b$ is a coupling constant. From the thermodynamical view, it is argued that the second law of thermodynamics strongly favors that dark energy decays into dark matter, which implies $b$ to be positive. These type of interactions are either positive or negative and can not change sign. However, recently by using a model independent method to deal with the observational data, Cai and Su found that the sign of interaction $Q$ in the dark sector changed in the redshift range of $0.45 \leq z \leq 0.9$.  Hereafter, a sign-changeable interaction \cite{Hao}-\cite{SChint1} were introduced,
\begin{equation}\label{eq:signcinteraction}
Q=q(\alpha\dot{\rho}+3\beta H\rho).
\end{equation}
where $\alpha$ and $\beta$ are dimensionless constants, the energy density $\rho$ could be $\rho_{m}$, $\rho_{\small{de}}$, $\rho_{tot}$. Also, $q$ is the deceleration parameter defined as,
\begin{equation}\label{eq:decparameter}
q=-\frac{1}{H^{2}} \frac{\ddot{a}}{a}=-1-\frac{\dot{H}}{H^{2}}.
\end{equation}
Deceleration parameter $q$ is a key ingredient makes this type of interactions different from the ones considered in literature and presented above, because it can change its sign when our Universe changes from deceleration $q>0$ to acceleration $q<0$. $\gamma \dot{\rho}$ is introduced from the dimensional point of view.\\
Dark energy models based on exotic nature of fluid and it is illustrated that, in nature, fluids with general form of EoS could be considered like to Chaplygin gas and its generalizations.
Subject of our interest is to consider two different models and study cosmological parameters. We consider composed models of an effective quintessence scalar field coupled with a,
\begin{enumerate}
\item barotropic fluid\\
\begin{equation}
P_{b}=\xi q\rho_{b},
\end{equation}
\item Viscous Polytropic gas
\begin{equation}
P_{VPG}=K\rho_{PG}^{1+\frac{1}{n}}-3\zeta H,
\end{equation}
\end{enumerate}
where $K$ and $n$ are the polytropic constant and polytropic index, respectively. Also, $q$ is a deceleration parameter given by the equation (\ref{eq:decparameter}).  The polytropic gas has some application in stellar astrophysics \cite{Christensen}.\\
The idea of dark energy with polytropic gas equation of state has been investigated by U. Mukhopadhyay and S. Ray in cosmology \cite{Mukhopadhyay}. Karami et al. \cite{Karami} investigated the interaction between dark energy and dark matter in polytropic gas scenario, the phantom behavior of polytropic gas, reconstruction of $f(T)$-gravity from the polytropic gas and the correspondence between polytropic gas and agegraphic dark energy model \cite{Karami}-\cite{Karami_2}. The cosmological implications of polytropic gas dark energy model is also discussed in \cite{Malekjani_1}.
The evolution of deceleration parameter in the context of polytropic gas dark energy model represents the decelerated expansion at the early Universe and accelerated phase later as expected. The polytropic gas model has also been studied from the viewpoint of statefinder analysis in \cite{Malekjani_2}. In addition to statefinder diagnostic, the other analysis to discriminate between dark energy models is $\omega-\omega^{\prime}$ analysis that have been used widely in the papers \cite{Khodam}-\cite{Huang1}.

\section{\large{The field equations}}
Field equations that govern our model of consideration are,
\begin{equation}\label{eq:Einstein eq}
R^{\mu\nu}-\frac{1}{2}g^{\mu\nu}R^{\alpha}_{\alpha}=T^{\mu\nu}.
\end{equation}
By using the
following FRW metric for a flat Universe,
\begin{equation}\label{s13}
ds^2=-dt^2+a(t)^2\left(dr^{2}+r^{2}d\Omega^{2}\right),
\end{equation}
field equations can be reduced to the following Friedmann equations,
\begin{equation}\label{eq: Fridmman vlambda}
H^{2}=\frac{\dot{a}^{2}}{a^{2}}=\frac{\rho}{3},
\end{equation}
and,
\begin{equation}\label{eq:Freidmann2}
\dot{H}=-\frac{1}{2}(\rho+P),
\end{equation}
where $d\Omega^{2}=d\theta^{2}+\sin^{2}\theta d\phi^{2}$, and $a(t)$
represents the scale factor.
Energy conservation $T^{;j}_{ij}=0$ reads as,
\begin{equation}\label{eq:Bianchi eq}
\dot{\rho}+3H(\rho+P)=0.
\end{equation}
The coupling between components may affect both the
background evolution of the Universe as well as the linear
growth of cosmological perturbations. Interaction between dark energy components may also plays an important role on small non-linear scales. Interacting dark energy models have also been considered as a possible solution to the coincidence problem.
In some unified models of dark energy the interaction between components may be strong enough for the dark sector as a
whole to be effectively described by a single fluid. However, in the cases of weak coupling, one can consider two-component fluid as dark matter with possibility of interaction between component. In order to introduce an interaction between DE and DM, one should split the equation (\ref{eq:Bianchi eq}) into the following equations [60],
\begin{equation}\label{eq:inteqm}
\dot{\rho}_{DM}+3H(\rho_{DM}+P_{DM})=Q,
\end{equation}
and,
\begin{equation}\label{eq:inteqG}
\dot{\rho}_{DE}+3H(\rho_{DE}+P_{DE})=-Q.
\end{equation}
where $Q$ represents an interaction term which can be an arbitrary function of cosmological parameters. Many candidates have been proposed in order to describe $Q$. Here, the interaction term is assumed to be of the following form,
\begin{equation}\label{eq:interQ}
Q=3Hb(\rho_{DE}+\rho_{DM}).
\end{equation}
An effective quintessence field will be characterized by a phenomenological assumption modifying energy density $\rho_{Q}$ and $P_{Q}$ pressure of the original field in the following way,
\begin{equation}\label{eq:qrho}
\rho_{Q}=\frac{\omega(t)}{2}\dot{\phi}^{2}+V(\phi),
\end{equation}
and
\begin{equation}\label{eq:qP}
P_{Q}=\frac{\omega(t)}{2}\dot{\phi}^{2}-V(\phi),
\end{equation}
where $\omega(t)$ obtained as,
\begin{equation}\label{eq:omegat}
\omega(t)=\omega_{0}+\omega_{1}(tH+t^{2}H^{2}).
\end{equation}
Analyzing general case numerically, when Universe is one component fluid expressed by an effective scalar field of our assumption we observed the following behavior, which is graphically indicated in Fig. 1. We can see that the Hubble expansion parameter is decreasing function of time and yields to a constant at the late time which is expected. Also, deceleration parameter yields to -0.5 as expected by observational data and shows acceleration to deceleration phase transition. Moreover the EoS is grater than -1 which is expected.

\begin{figure}[h!]
 \begin{center}$
 \begin{array}{cccc}
\includegraphics[width=50 mm]{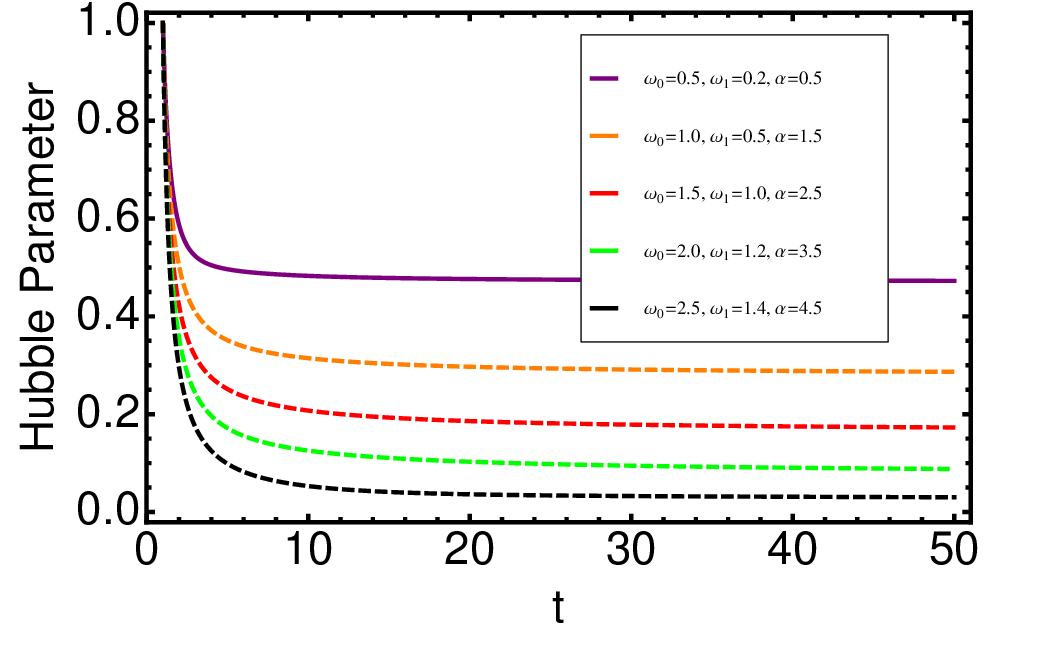} &
\includegraphics[width=50 mm]{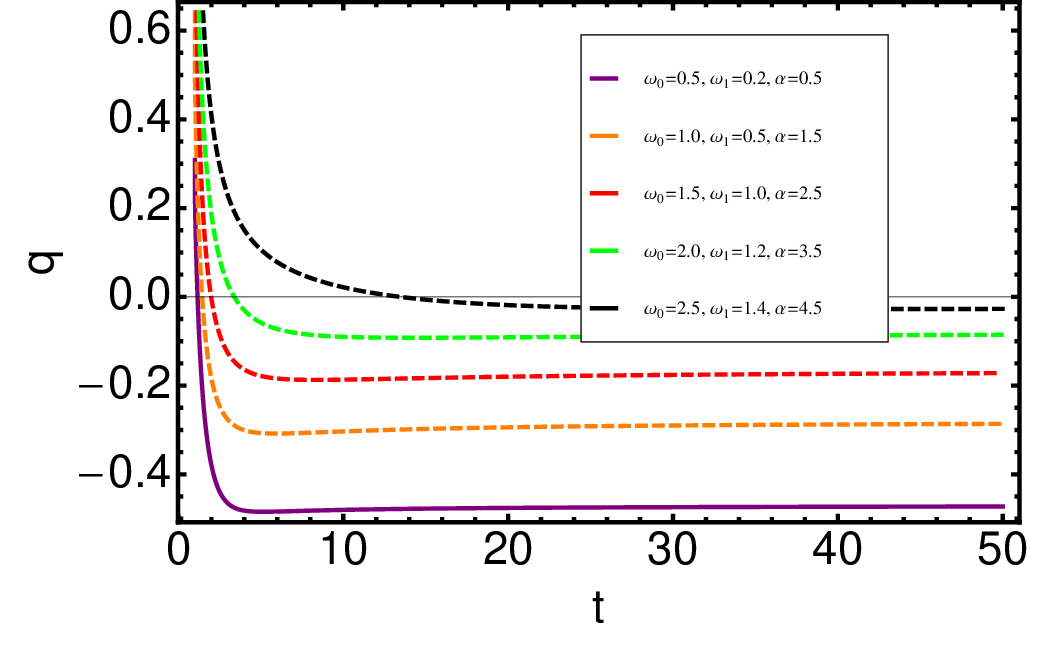}&
\includegraphics[width=50 mm]{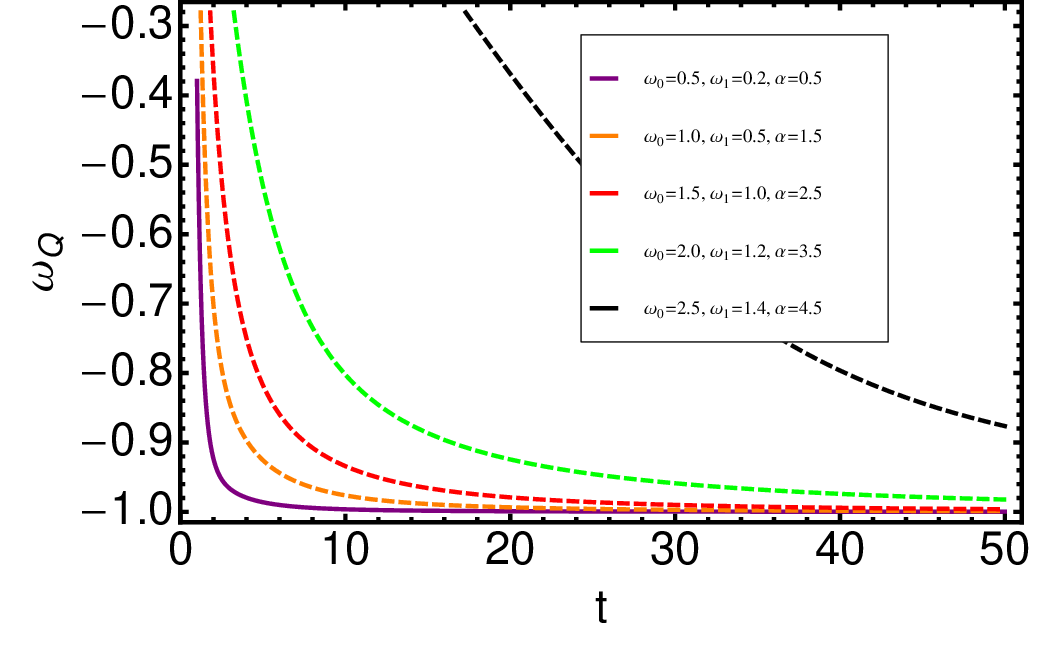} &
 \end{array}$
 \end{center}
\caption{Behavior of Hubble $H$ and deceleration parameter $q$ and $\omega_{Q}=P_{Q}/\rho_{Q}$ against  cosmic $t$. Case corresponds to a Universe with an effective quintessence scalar field.}
 \label{fig:1}
\end{figure}
$H(z)$ is presented in Fig. 2 in agreement with observational data [62, 63]. The best fit with observations provided by $\omega_{0}=1.0$, $\omega_{1}=0.5$, $H_{0}=1.4$ and $\alpha=1.2$. Generally we will see that for all models the typical value of $\alpha$ is between $1$ and $4$.

\begin{figure}[h!]
 \begin{center}$
 \begin{array}{cccc}
\includegraphics[width=50 mm]{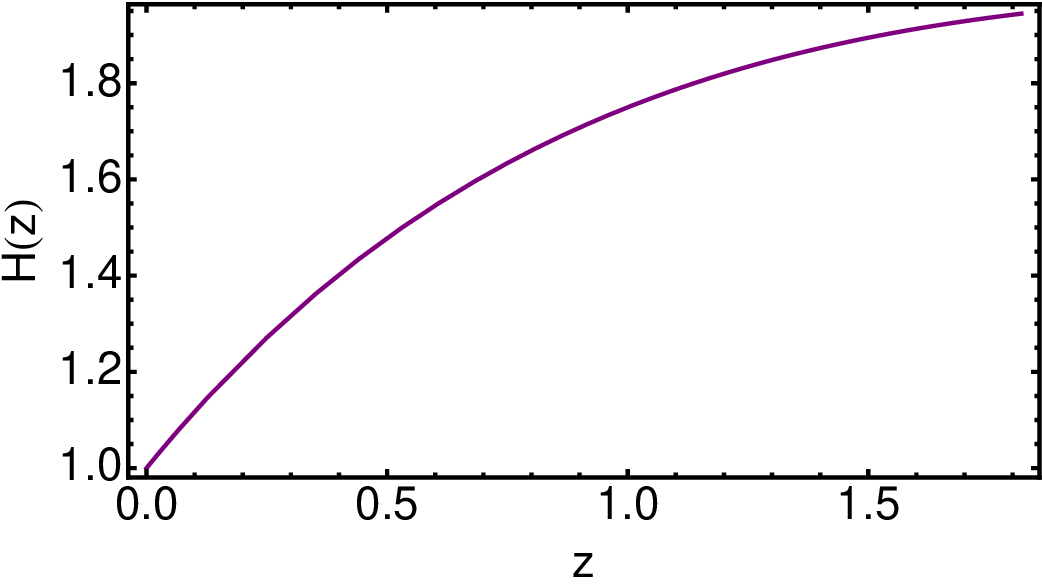}
 \end{array}$
 \end{center}
\caption{ Plot of $H(z)$. Case corresponds to a Universe with an effective quintessence scalar field.}
 \label{fig:2}
\end{figure}

For a potential of the scalar field we assume,
\begin{equation}\label{eq:potential}
V(\phi)=\phi^{-\alpha}.
\end{equation}
In next two sections we will analyze two-component interacting Universe as well as case corresponding to a Universe with a barotropic fluid which EoS parameter is a function of deceleration parameter $q$. We will consider a simple case, where $\omega_{b}=\xi q$ is a linear function of $q$.
\section{\large{Universe with a barotropic fliud with $\omega_{b}=\xi q$ EoS parameter}}
Before consider interacting models and behavior of the Universe, we would like to consider a behavior of the Universe within our assumption, where we have associated EoS parameter of a barotropic fluid with the deceleration parameter as a linear function. In this section, similar the case of effective scalar field, which is the second phenomenological assumption of our work, we will analyze model numerically and recover graphical behavior of Hubble parameter, deceleration parameter and so forth as a function of a $\xi$ single parameter of the model (see Fig. 3). The best fit was possible to provide with $\xi=3.5$. As we can see the Hubble parameter behaves normal as previous case, but we see new behavior of the deceleration parameter. In the cases of small $\xi$ we obtain totally positive $q$ which vanishes at the late time. On the other hand larger $\xi$ may gives expected value of $q$ at the late time, however no acceleration to deceleration phase transition exist in this model which is inconsistent with observational data.

\begin{figure}[h!]
 \begin{center}$
 \begin{array}{cccc}
\includegraphics[width=50 mm]{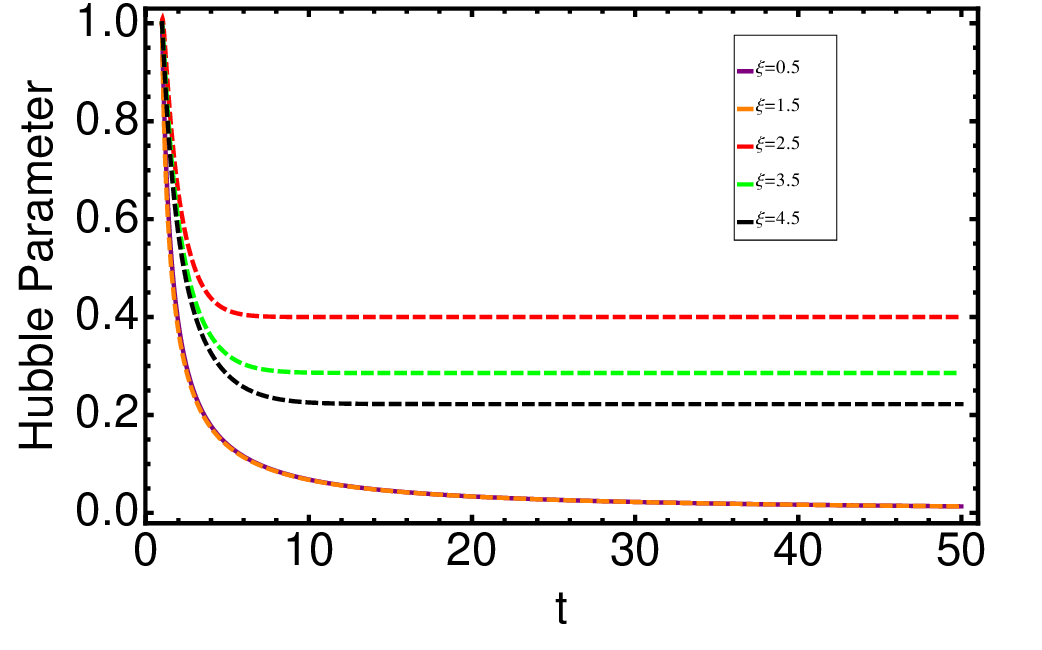} &
\includegraphics[width=50 mm]{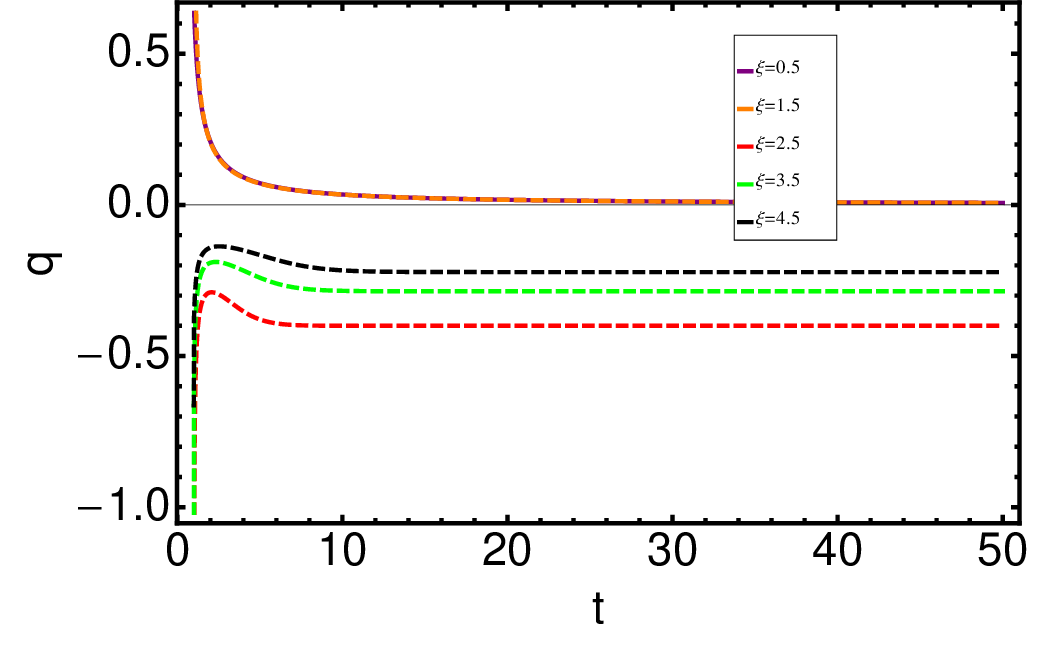}
 \end{array}$
 \end{center}
\caption{Behavior of Hubble parameter $H$ and $q$ against cosmic time $t$ as functions of $\xi$. $\xi=0.5,1.5,2.5,3.5,4.5$. Case describes the Universe with a barotropic fluid.}
 \label{fig:3}
\end{figure}

\section{\large{Two component interacting models}}
In this section we consider two different models interacting with the quintessence dark energy. Cosmological parameters of our interest are EoS parameters of each fluid components,
\begin{equation}\label{24}
\omega_{i}=P_{i}/\rho_{i},
\end{equation}
and EoS parameter of composed fluid,
\begin{equation}\label{25}
\omega_{tot}=\frac{P_{i}+P_{Q} }{\rho_{i}+\rho_{Q}},
\end{equation}
where index $i$ refers to barotropic fluid by $b$, and polytropic gas by $PG$.
Deceleration parameter $q$ written as,
\begin{equation}\label{eq:accchange}
q=\frac{1}{2}(1+3\frac{P}{\rho} ),
\end{equation}
\subsection{\large{Model 1: Interacting barotropic fluid and scalar field}}
The model of interacting barotropic fluid $P_{b}=\xi q \rho_{b}$ with scalar field we can describe with a total energy density and pressure like $\rho=\rho_{b}+\rho_{Q}$ and $P=P_{b}+P_{Q}$. Dynamics of energy density of a barotropic fluid will take following form after taking into account our assumptions,
\begin{equation}\label{eq:bfluid}
\dot{\rho}_{b}+3H(1+\xi q-b)\rho_{b}=3Hb\rho_{Q}.
\end{equation}
Dynamics of energy density of the scalar field can be written as,
\begin{equation}\label{eq:scalarfield}
\frac{\dot{\omega(t)}}{2}\dot{\phi}+\omega(t)\left( \ddot{\phi}+\frac{3}{2}Hb + 3H\dot{\phi}\right) - \phi^{-\alpha} (\alpha\phi^{-1}-\frac{3Hb}{\dot{\phi}})=3Hb\frac{\rho_b}{\dot{\phi}}.
\end{equation}
Behavior of total energy density and pressure is also investigated numerically for different values of parameters of the model which can find in in Fig. 4 together with graphical behavior of Hubble parameter and deceleration parameter $q$. We can see expected behavior for the Hubble expansion parameter. Deceleration parameter begin with positive value and yields to -1 at the late time in agreement with $\Lambda$CDM model. Therefore we can see acceleration to deceleration phase transition. Also we can see that total energy density is decreasing function of time and yields to infinitesimal value at the late time which expected. Total pressure has positive value at the early Universe and has negative value at the late Universe which is also expected.

\begin{figure}[h!]
 \begin{center}$
 \begin{array}{cccc}
\includegraphics[width=50 mm]{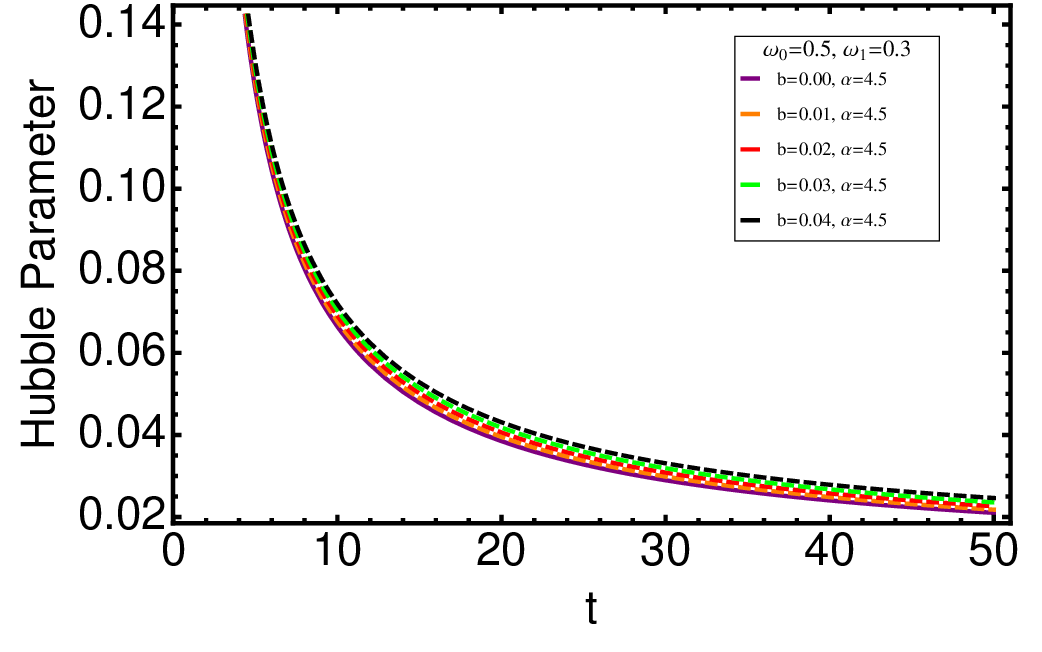}
\includegraphics[width=50 mm]{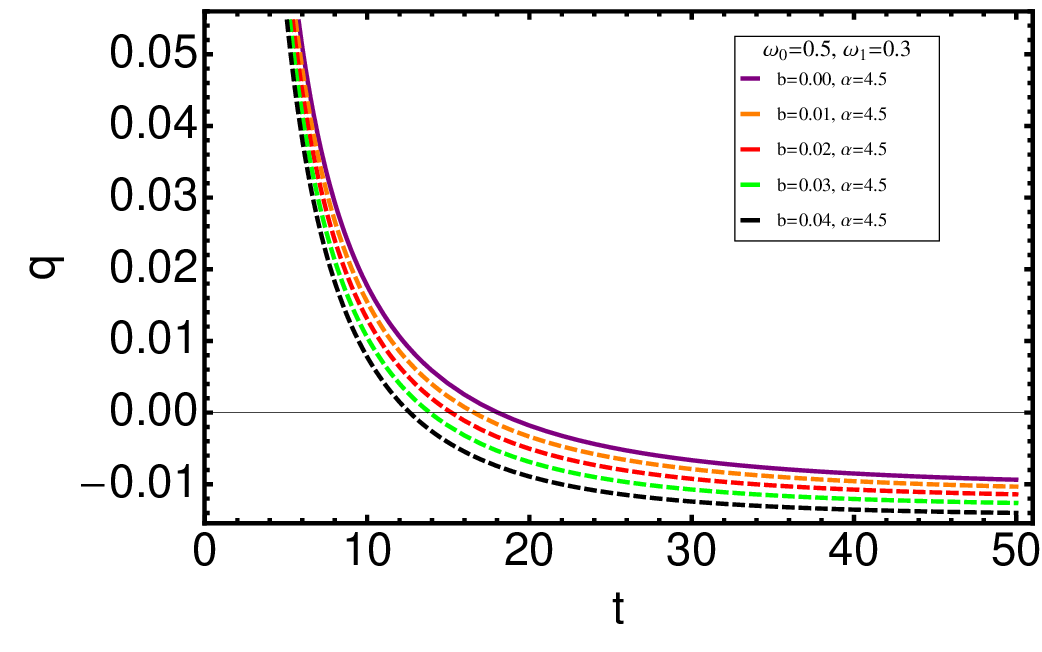}\\
\includegraphics[width=50 mm]{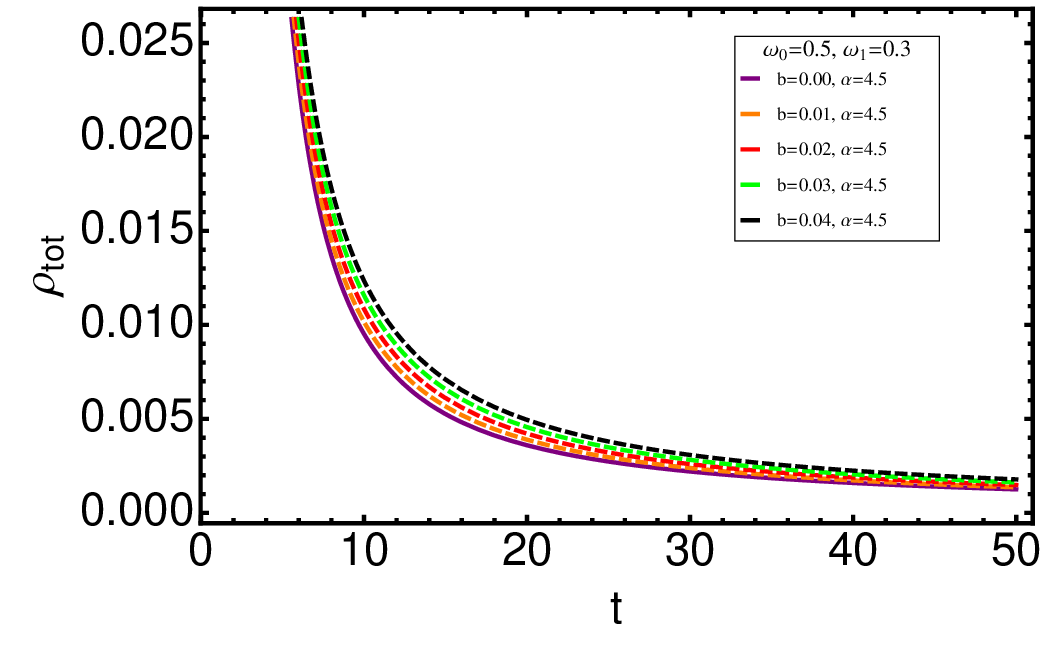}
\includegraphics[width=50 mm]{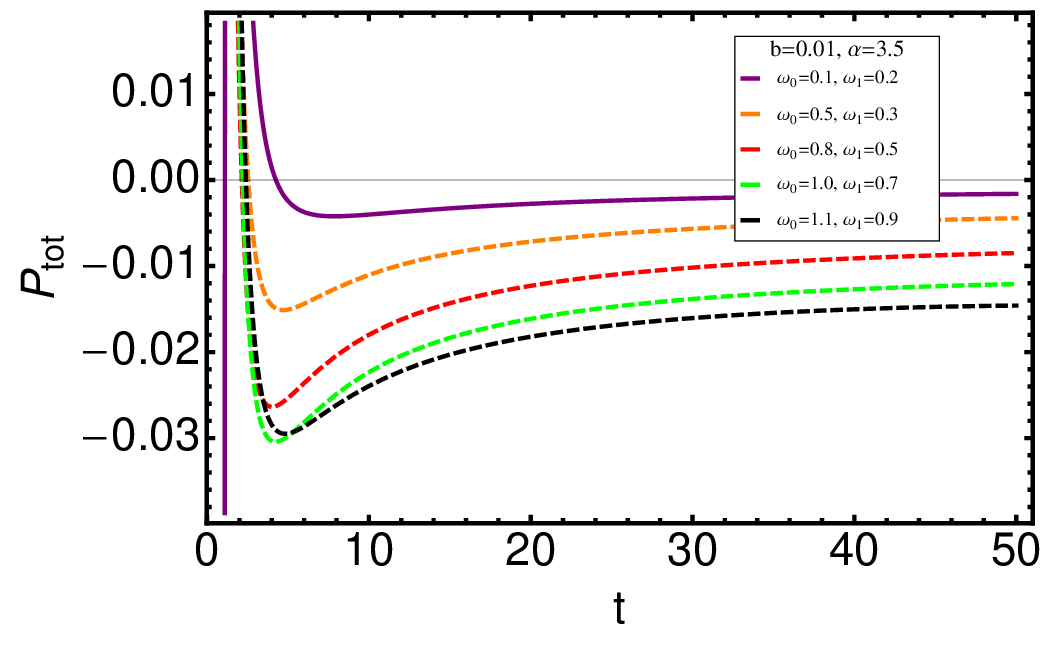}
 \end{array}$
 \end{center}
\caption{Behavior of Hubble $H$, deceleration parameter $q$, total energy density and total pressure against $t$. Case corresponds to a Universe with an interacting quintessence scalar field and barotropic fluid.}
 \label{fig:4}
\end{figure}
Behavior of field $\phi$ against time given in Fig. 5, which shows quintessence field is increasing function of time and yields to a constant at the late time. Fig. 6 is to present behavior of $\omega_{Q}$ and $\omega_{tot}$, we can see that both of them yields to -1 at the late time.
\begin{figure}[h!]
 \begin{center}$
 \begin{array}{cccc}
\includegraphics[width=50 mm]{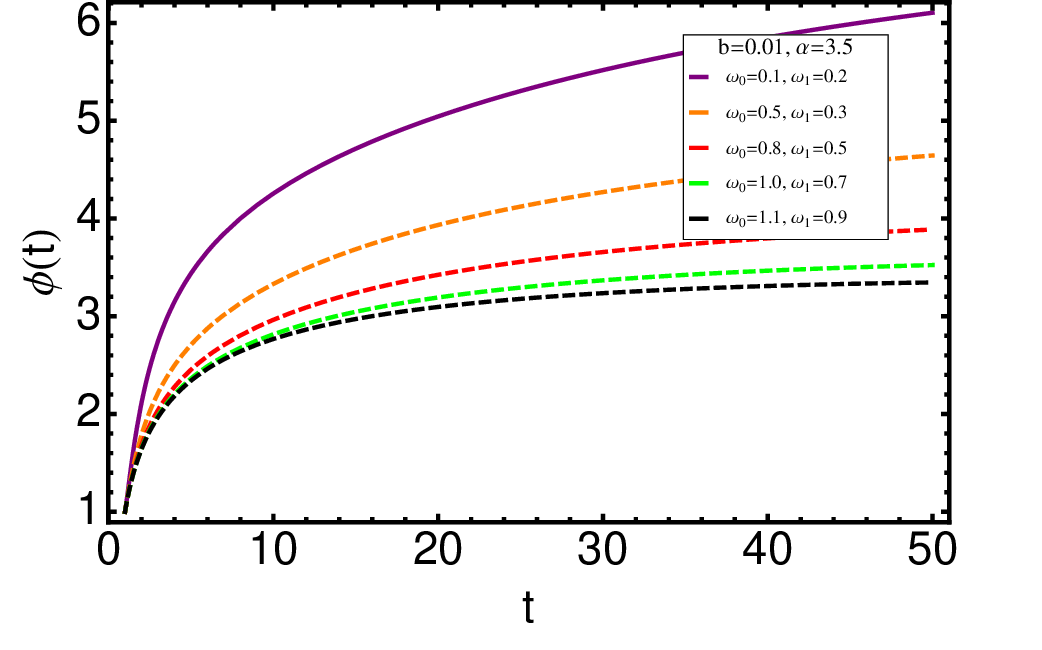}
 \end{array}$
 \end{center}
\caption{Behavior of the scalar field $\phi$ against $t$. Case corresponds to a Universe with an interacting effective quintessence scalar field and barotropic fluid.}
 \label{fig:5}
\end{figure}

\begin{figure}[h!]
 \begin{center}$
 \begin{array}{cccc}
\includegraphics[width=50 mm]{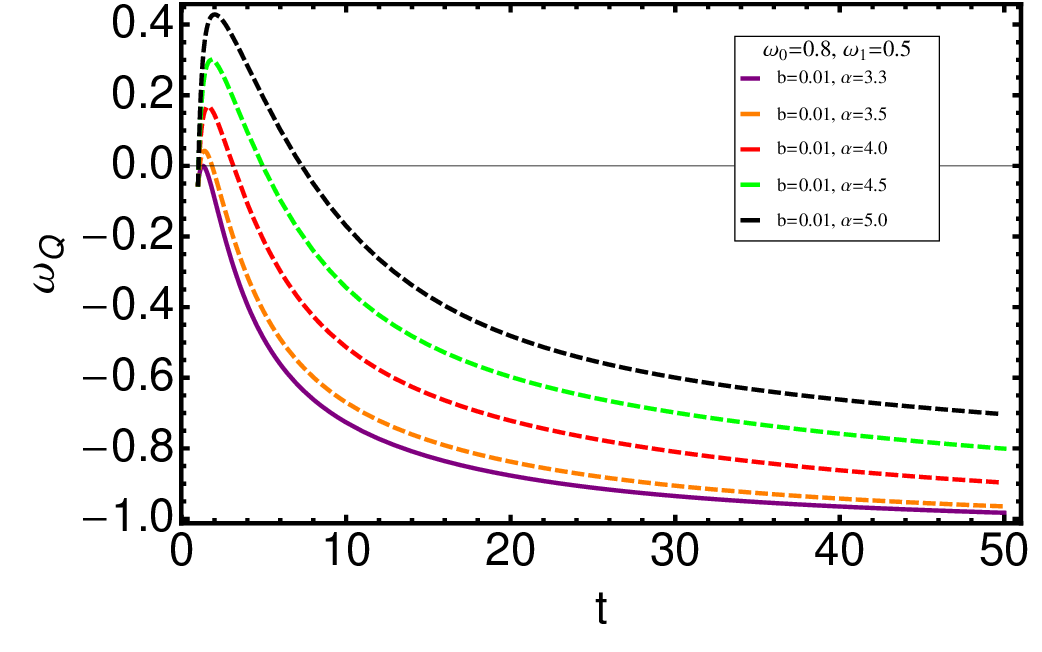} &
\includegraphics[width=50 mm]{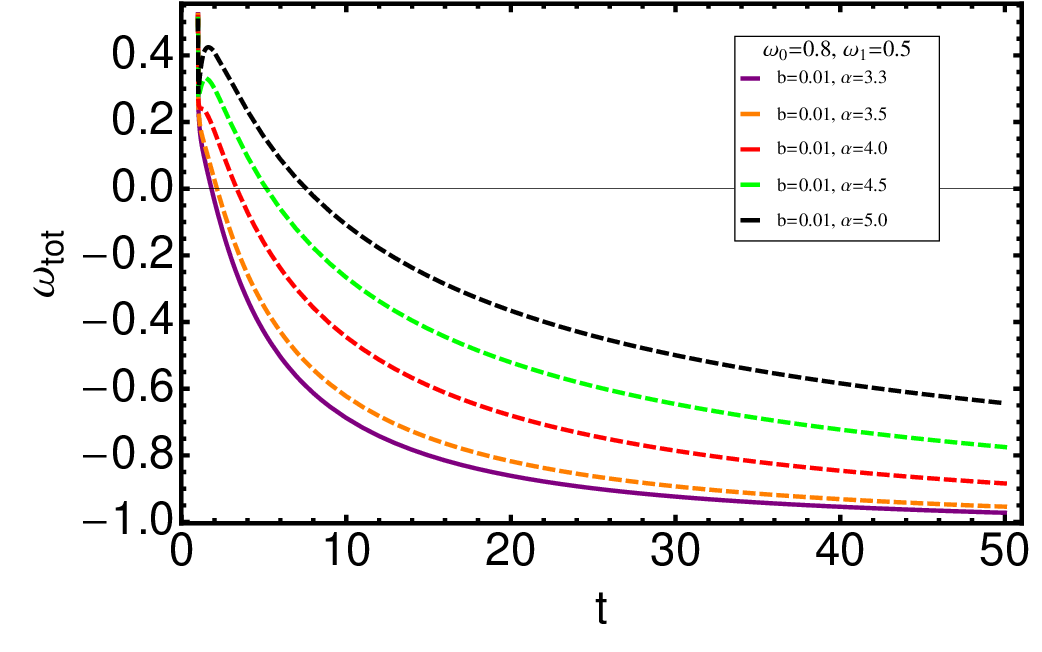} &
 \end{array}$
 \end{center}
\caption{Behavior of $\omega_{q}$ and $\omega_{tot}$ against $t$. Case corresponds to a Universe with an interacting effective quintessence scalar field and barotropic fluid.}
 \label{fig:6}
\end{figure}

\begin{figure}[h!]
 \begin{center}$
 \begin{array}{cccc}
\includegraphics[width=50 mm]{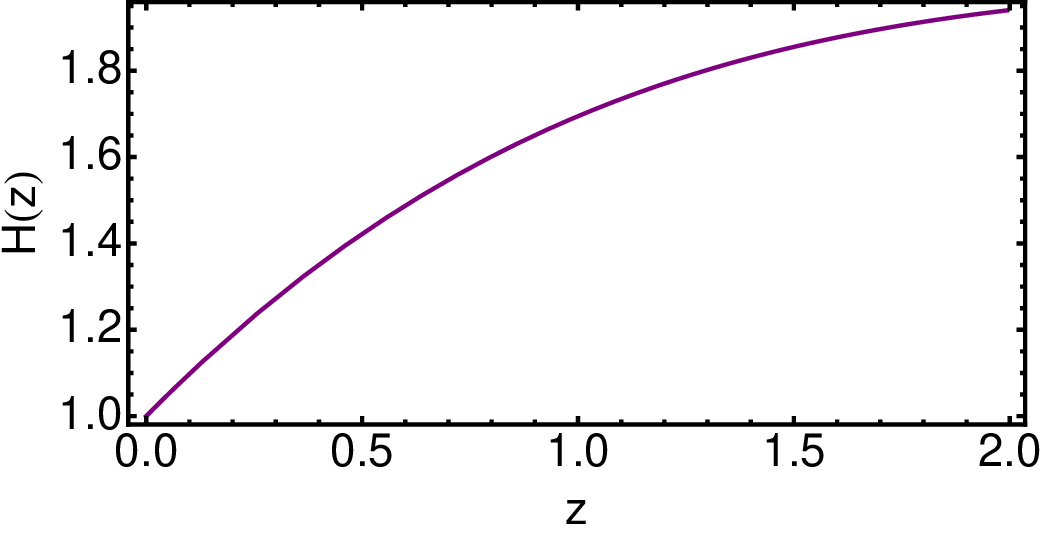} &
 \end{array}$
 \end{center}
\caption{Behavior of $\omega_{q}$ and $\omega_{tot}$ against $t$. Case corresponds to a Universe with an interacting effective quintessence scalar field and barotropic fluid.}
 \label{fig:6}
\end{figure}

\begin{figure}[h!]
 \begin{center}$
 \begin{array}{cccc}
\includegraphics[width=50 mm]{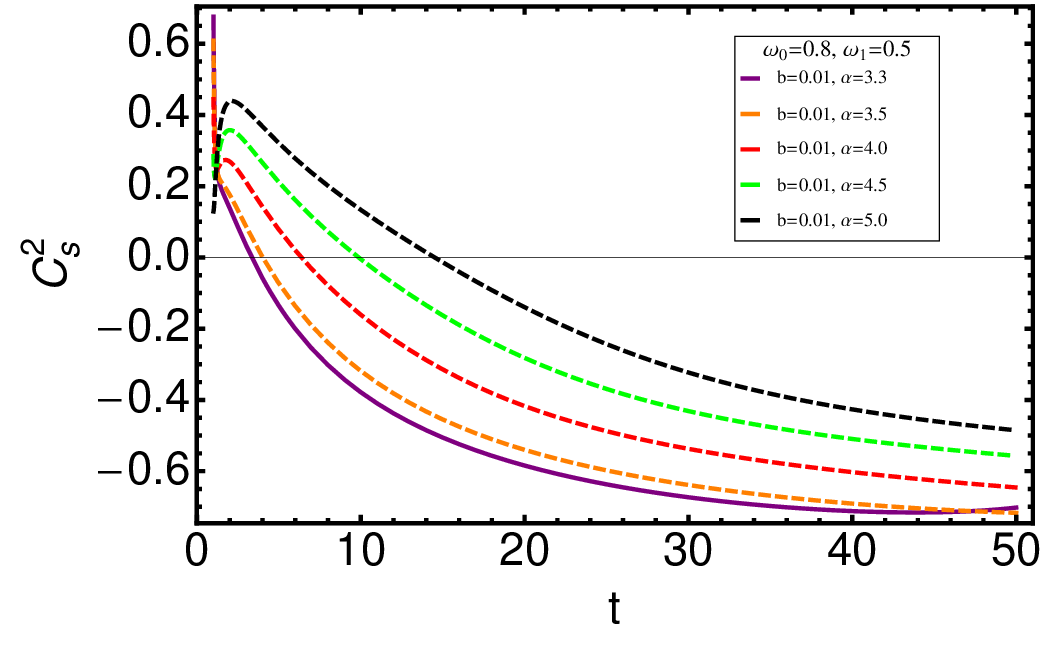}&
\includegraphics[width=50 mm]{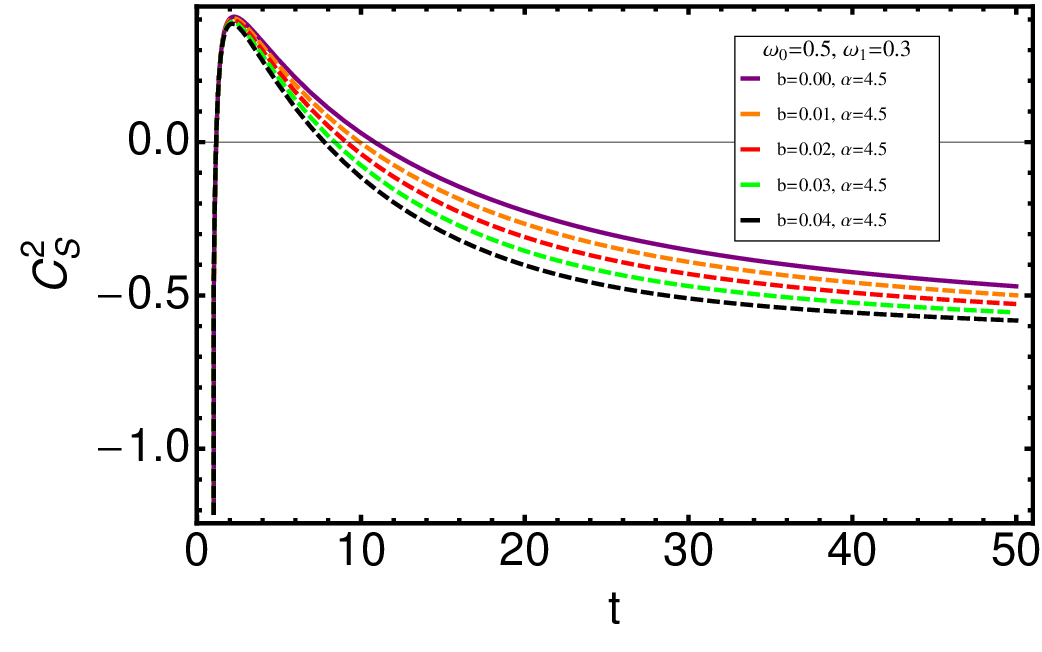}
 \end{array}$
 \end{center}
\caption{Squared sound speed against $t$. Case corresponds to a Universe with an interacting effective quintessence scalar field and barotropic fluid.}
 \label{fig:muz}
\end{figure}
Behavior of $H(z)$ also agree with observational data (see Fig. 7). Finally analysis of squared sound speed shows that this model is stable at the early Universe. It may be there is stability at present but our model is instable at the late time (see Fig. 8).
\subsection{\large{Model 2: Interacting viscous polytropic gas and scalar field}}
Taking into account EoS equation for a vicious polytropic gas dynamics of its energy density can be written as,
\begin{equation}
\dot{\rho}_{PG}+3H(1+K\rho_{PG}^{\frac{1}{n}}+b)\rho_{PG}=-3Hb\rho_{Q}+9H^{2}\zeta.
\end{equation}

\begin{figure}[h!]
 \begin{center}$
 \begin{array}{cccc}
\includegraphics[width=50 mm]{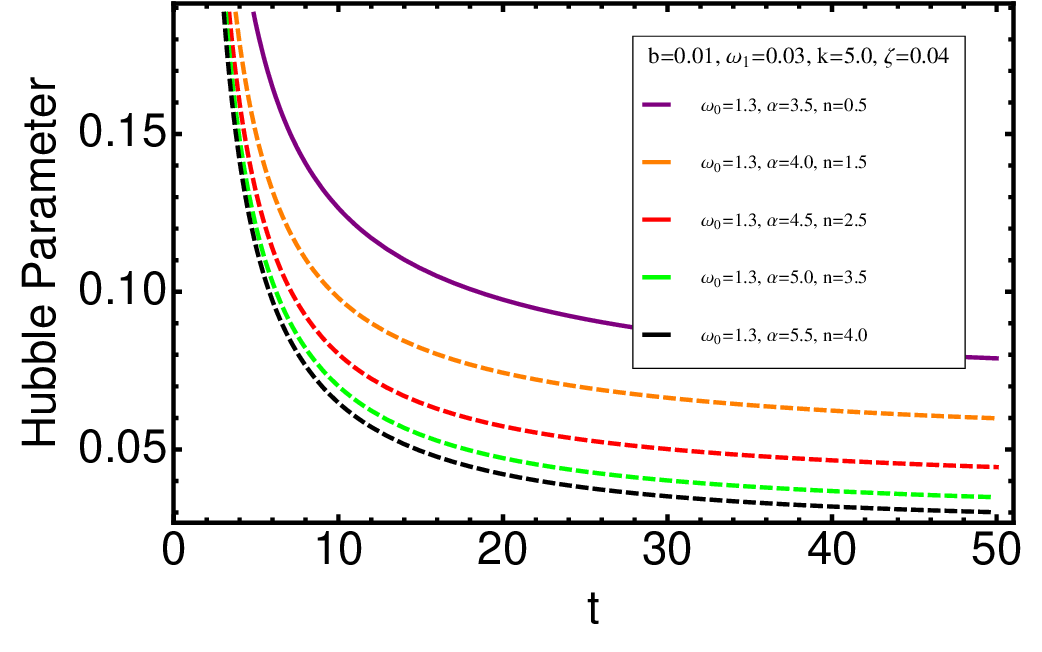}
\includegraphics[width=50 mm]{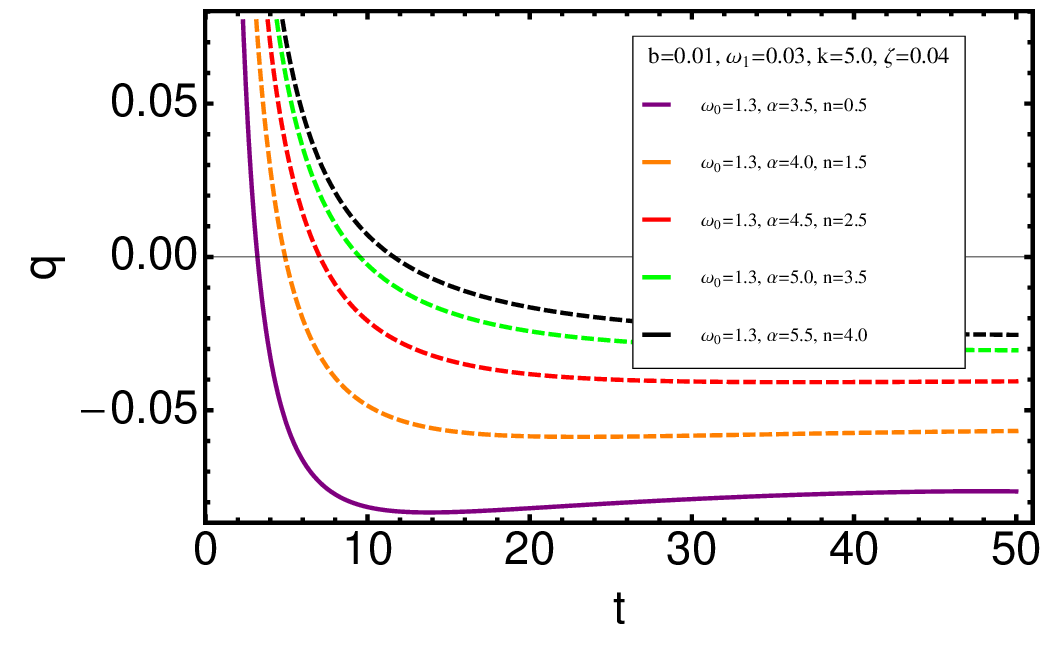}
 \end{array}$
 \end{center}
\caption{Behavior of Hubble $H$ and deceleration parameter $q$ against $t$. Case corresponds to a Universe with an interacting effective quintessence scalar field and vicious polytropic gas.}
 \label{fig:7}
\end{figure}

\begin{figure}[h!]
 \begin{center}$
 \begin{array}{cccc}
\includegraphics[width=50 mm]{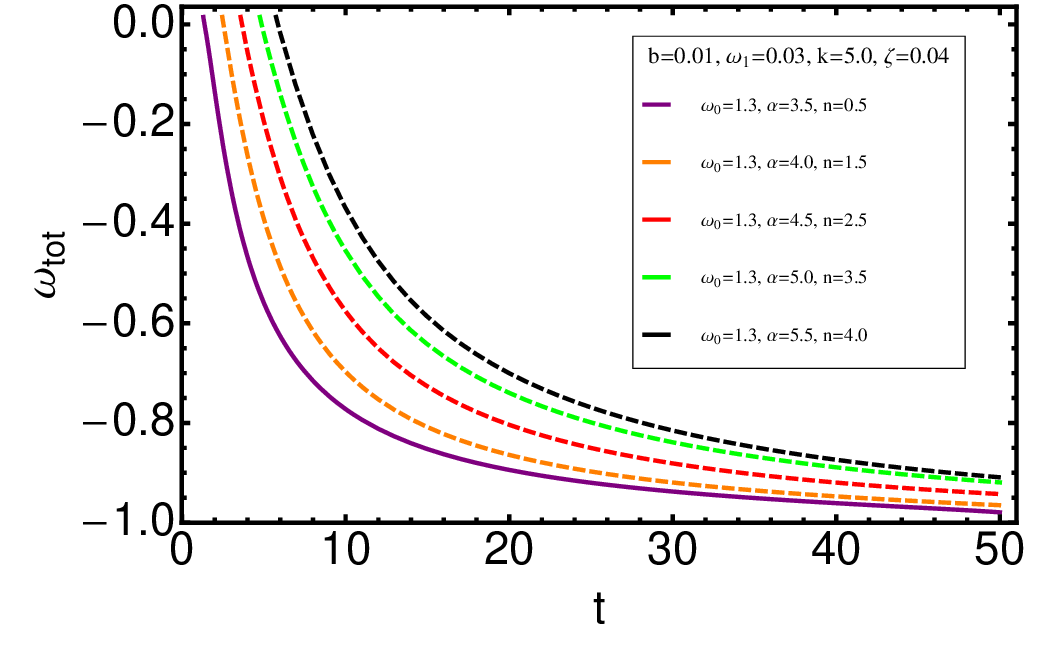}
 \end{array}$
 \end{center}
\caption{Behavior of $\omega_{tot}$ against $t$. Case corresponds to a Universe with an interacting effective quintessence scalar field and vicious polytropic gas.}
 \label{fig:9}
\end{figure}

For this model we investigate cosmological parameters for a general case and numerically recover behavior of Hubble parameter, field, deceleration parameter and so forth. Graphical behavior of Hubble parameter and deceleration parameter can be found in Fig. 9. As before we can see that the Hubble parameter has usual behavior and $q$ yields to about -0.5 consistent with some observational data. Also acceleration to deceleration phase transition exists in this model.\\
Behavior of $\omega_{tot}$ against time is given in Fig. 10. It is illustrated that $\omega_{tot}\rightarrow-1$ at the late time, but at this time our model has some instabilities which presented in Fig. 11. However, by choosing appropriate values of parameter we can construct stable model at the late time.

\begin{figure}[h!]
 \begin{center}$
 \begin{array}{cccc}
\includegraphics[width=50 mm]{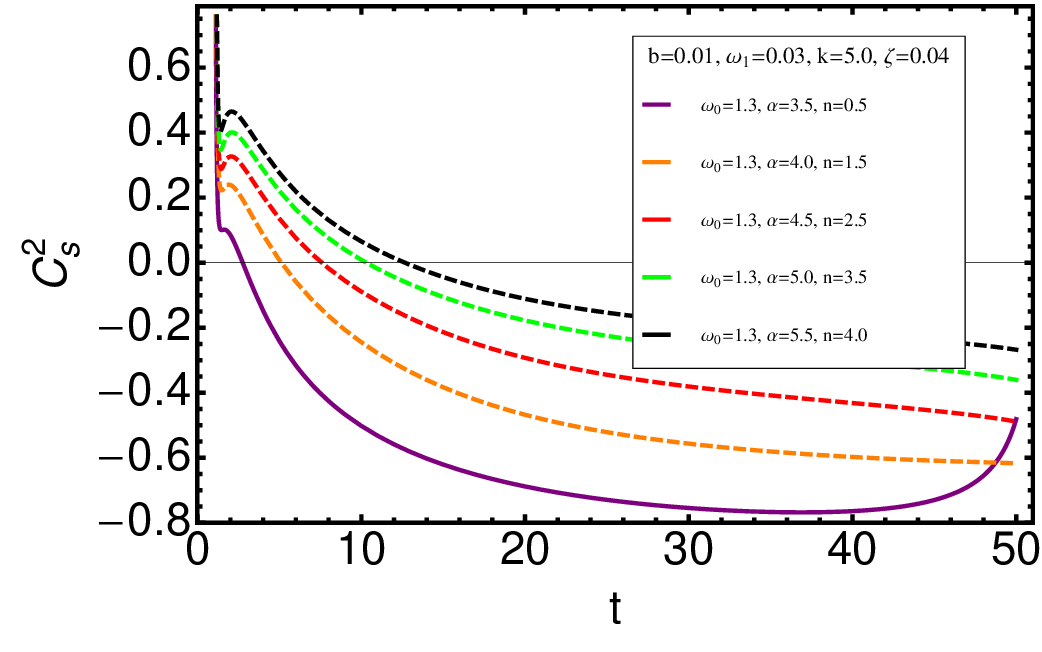}&
\includegraphics[width=50 mm]{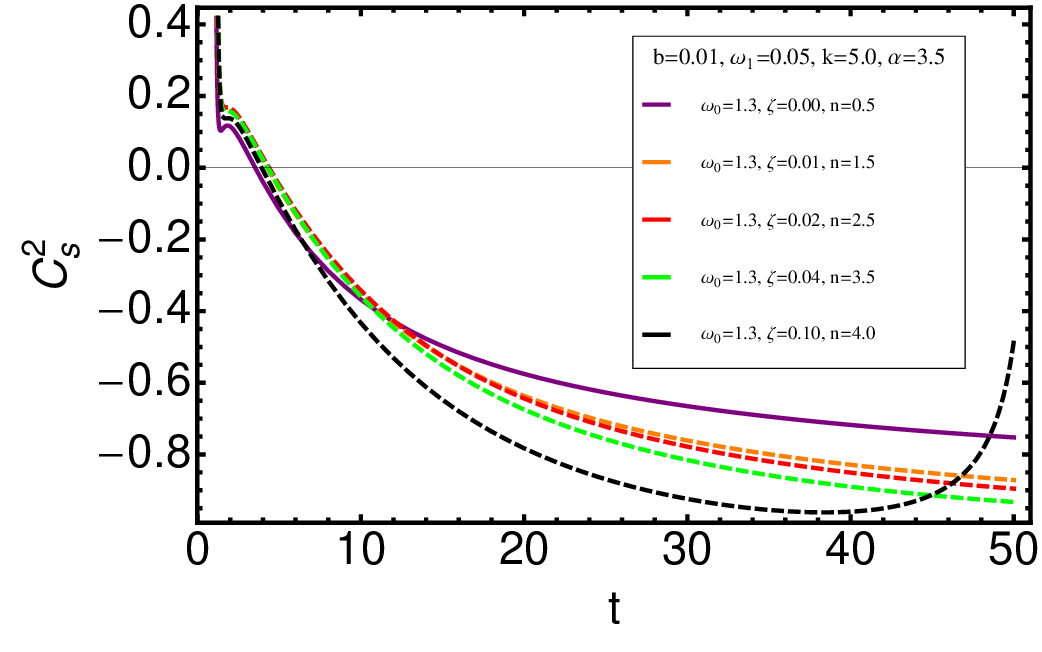}
 \end{array}$
 \end{center}
\caption{Squared sound speed against $t$. Case corresponds to a Universe with an interacting effective quintessence scalar field and vicious polytropic gas.}
 \label{fig:muz}
\end{figure}
\section{Observational constraints on interacting models}
In this section we use recent observational data to fix parameter of the models in the case of interacting with the quintessence dark energy.
As we know the SNIa observations is based on the distance modulus $\mu$ given by [64],
\begin{equation}
\mu=m-M=5\log_{10}{D_L},
\end{equation}
where $D_{L}$ is the luminosity
distance defined as,
\begin{equation}
D_{L}=(1+z)\frac{c}{H_{0}}\int_{0}^{z}{\frac{dz'}{\sqrt{H(z')}}}.
\end{equation}
The quantities $m$ and $M$ denote the apparent and the absolute magnitudes, respectively. There are several SNIa data sets, which are obtained by using different techniques which may yield to very different results. Our observational analysis of the background dynamics includes the differential age of old objects as well as the data of SNIa, BAO and CMB. We used mentioned observational data to fix parameters of both models which presented in the following table.\\

\begin{center}
    \begin{tabular}{ | l | l | l | l | l | l | l |l | l | l | }
    \hline
    M & $\omega_{1}$ & $\omega_{0}$  & $\xi$  &  $\alpha$ & $K$ & $n$ & $b$ & $H_{0}$ & $\Omega_{m0}$ \\ \hline
    1 & $0.05^{+0.01}_{-0.01}$ & $1.05^{+0.05}_{-0.05}$ & $0.05^{+0.02}_{-0.02}$ & $1.2^{+0.6}_{-0.2}$ & $3.0^{+1.0}_{-1.0}$ &  $1.5^{+0.75}_{-0.25}$ &  $0.01^{+0.01}_{-0.01}$ &  $1.5^{+0.1}_{-0.1}$ &  $0.35^{+0.05}_{-0.15}$\\ \hline
    2 & $0.5^{+0.3}_{-0.2}$ & $0.5^{+0.25}_{-0.15}$ & $-$ & $1.1^{+0.1}_{-0.1}$ & $-$ &  $-$ &  $0.01^{+0.01}_{-0.01}$ &  $1.65^{+0.05}_{-0.1}$ &  $0.43^{+0.02}_{-0.1}$ \\ \hline
    \end{tabular}
\end{center}

In the Figs. 12, 13 and 14 we can see and compare parameter $\mu$ with observational data obtained by combination of SneIa+BAO+CMB. In all cases we can see good agreement with observations.\\

\begin{figure}[h!]
 \begin{center}$
 \begin{array}{cccc}
\includegraphics[width=50 mm]{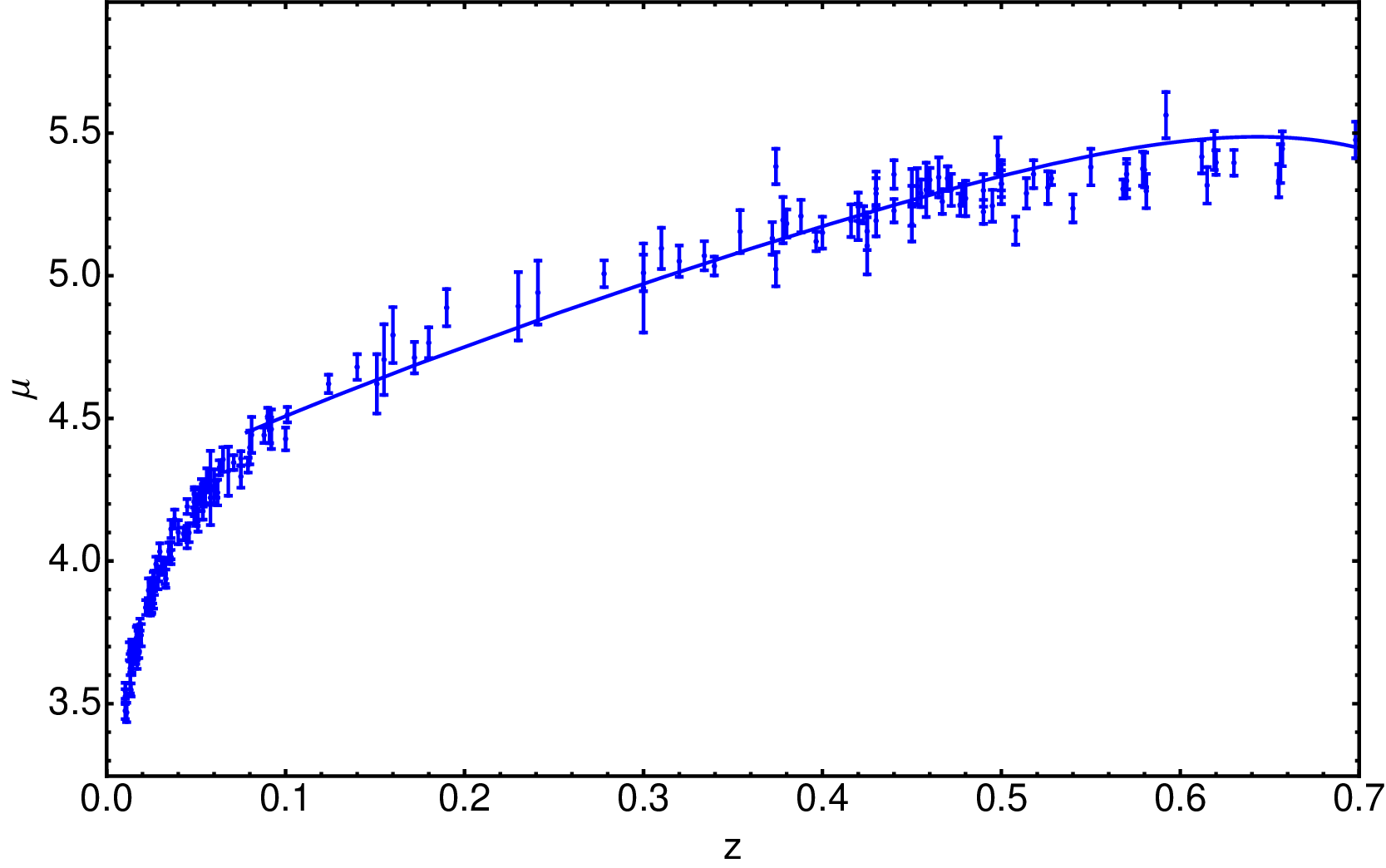}
 \end{array}$
 \end{center}
\caption{ Observational data  SneIa+BAO+CMB for distance modulus versus redshift $z$. Case corresponds to a Universe with an effective quintessence scalar field.}
 \label{fig:2}
\end{figure}

\begin{figure}[h!]
 \begin{center}$
 \begin{array}{cccc}
\includegraphics[width=50 mm]{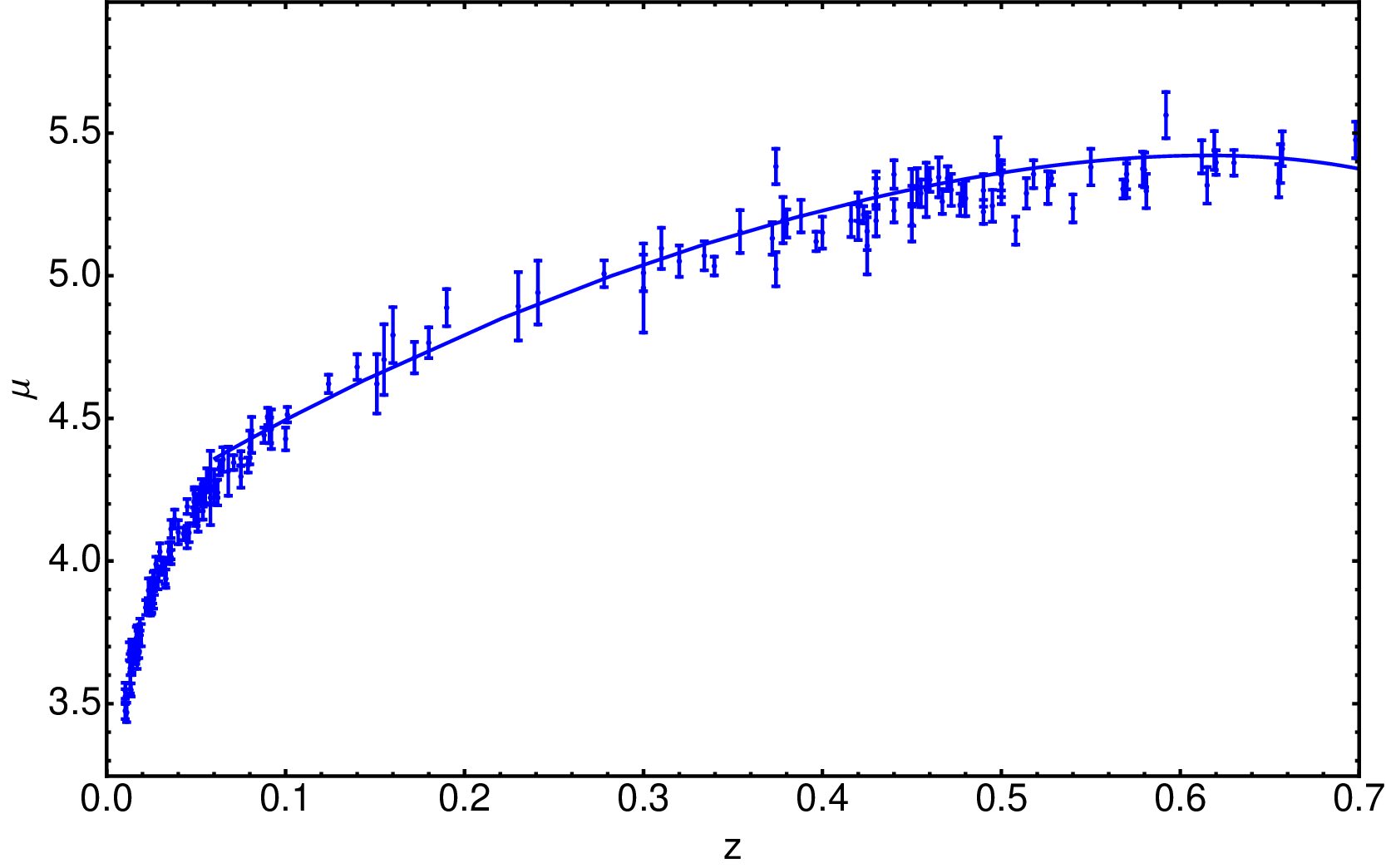}
 \end{array}$
 \end{center}
\caption{ Observational data SneIa+BAO+CMB for distance modulus versus redshift $z$. Case describes the Universe with a barotropic fluid.}
 \label{fig:4}
\end{figure}

\begin{figure}[h!]
 \begin{center}$
 \begin{array}{cccc}
\includegraphics[width=50 mm]{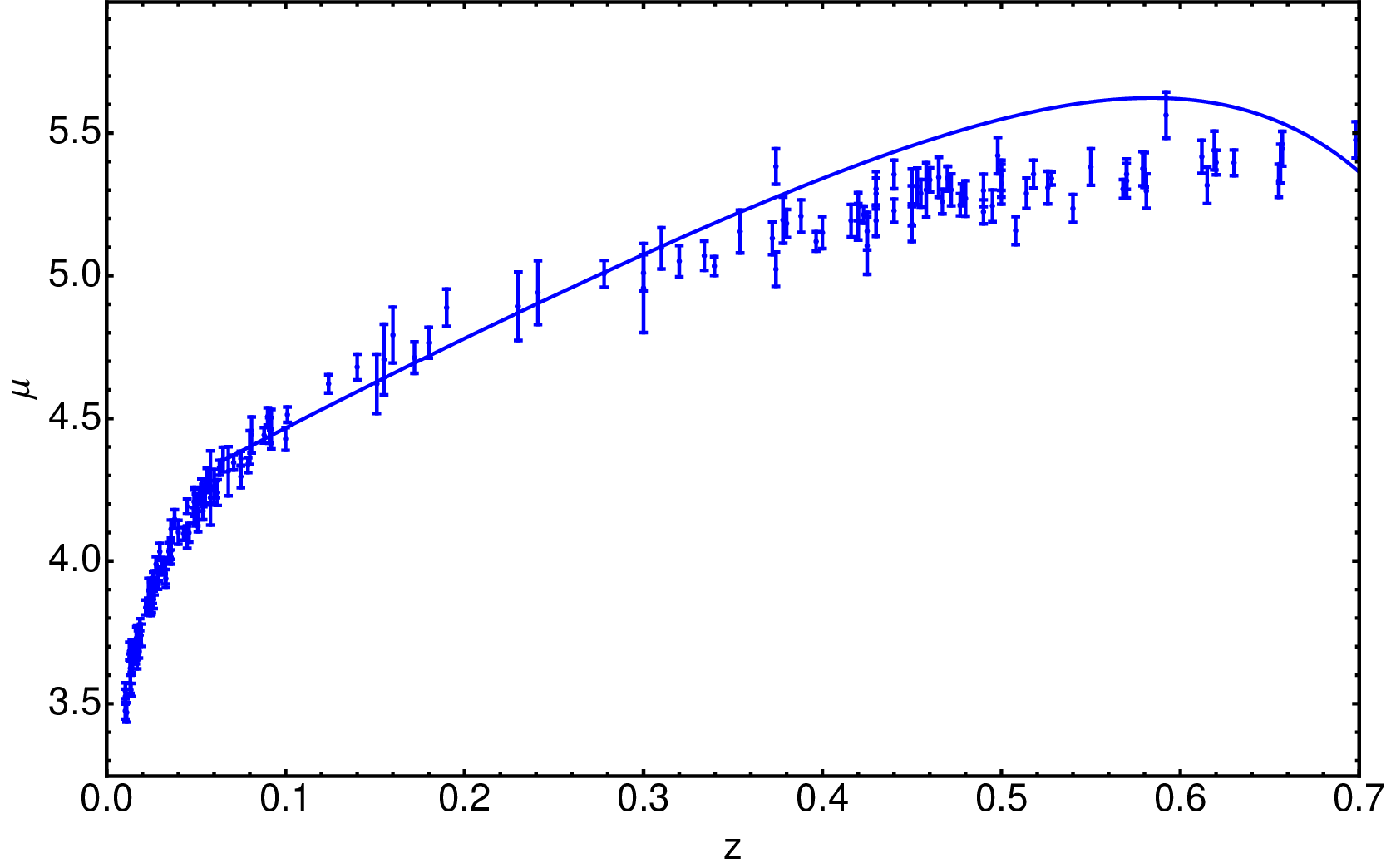}&
\includegraphics[width=50 mm]{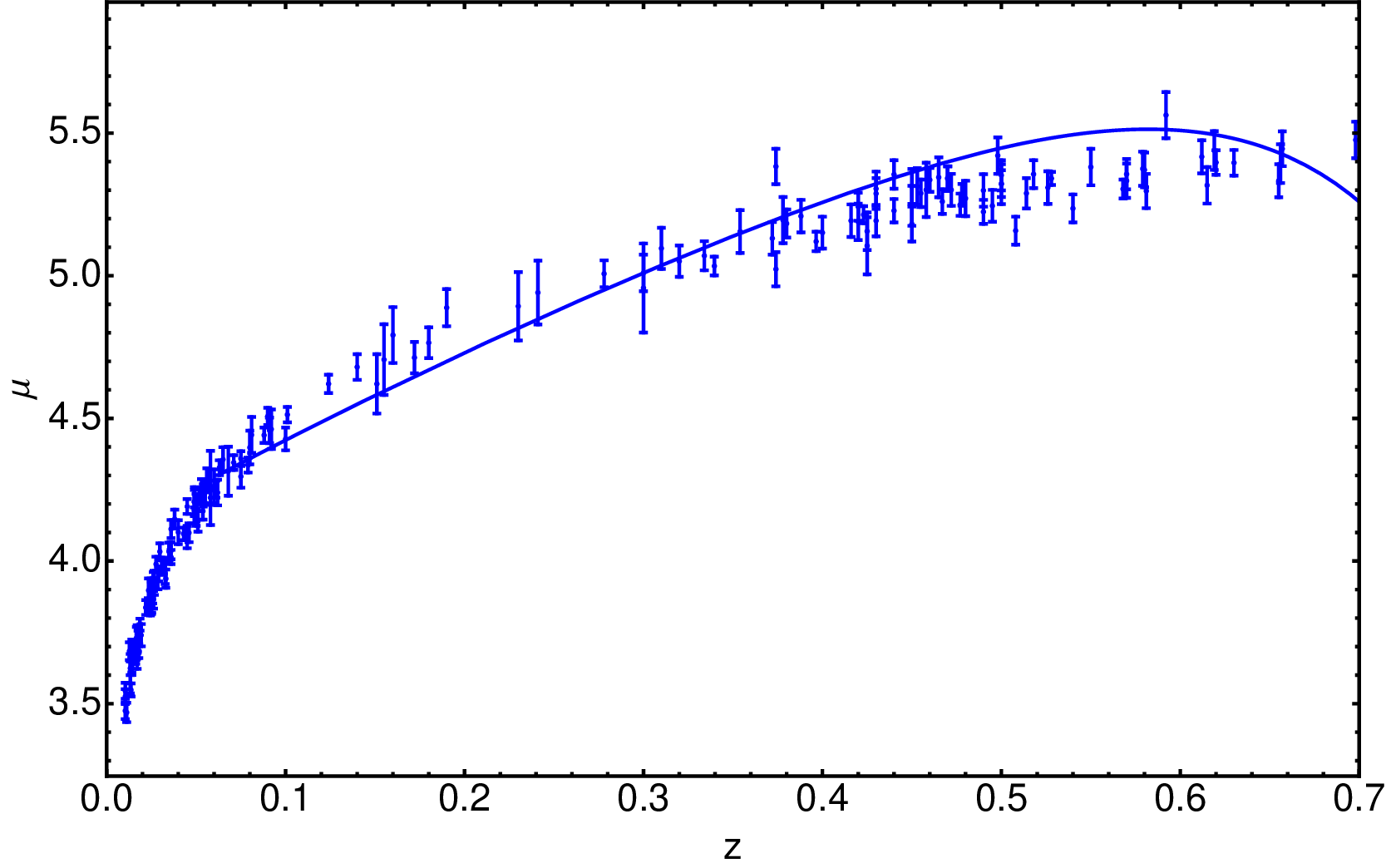}
 \end{array}$
 \end{center}
\caption{Observational data SneIa+BAO+CMB for distance modulus versus redshift $z$. Case of interacting with the quintessence dark energy in models 1 and 2 respectively.}
 \label{fig:muz}
\end{figure}

\section{Discussion}
In this paper, we considered an effective quintessence cosmology with possibility of interaction between other components. Indeed, we considered a two-component fluid to describe Universe. The first component assumed quintessence dark energy, and we have two possibilities for the second component: barotropic fluid or viscous Polytropic gas. First of all we discussed about single-component Universe. Then, we added interaction term and considered two-component fluid as a model for the Universe. We analyzed numerically about cosmological parameters and found that $\omega\geq-1$ verified in agreement with quintessence cosmology. We found some instabilities at the late time in all models, but by choosing appropriate parameters in the model of viscous Polytropic gas interacting with scalar field the instabilities may removed. However, always we can have stable models at the early Universe. Hubble parameter has similar manner for the cases of single-component, but its value decreased dramatically in the cases of interacting two-component. Behavior of Hubble parameter in terms of redshift shows agreement of our models with some observational data, however further investigations need to confirm validity of these models. In order to have more suitable models we fixed our parameters by using observational data of SneIa+BAO+CMB which summarized in a table.\\
In summary we can propose interacting quintessence with the viscous Polytropic gas as a toy model for the Universe which has good agreement with observational data. This model may be considered together with various successful models of dark energy waiting for more observational data to confirmation.

\end{document}